\documentclass[aps,twocolumn,showpacs,amsmath,amssymb,nofootinbib, prc]{revtex4-1}

\usepackage{graphicx}
\usepackage{amsmath}
\usepackage{hyperref}
\usepackage{footmisc}
\usepackage{chngcntr}

\hypersetup{
	colorlinks=true,
	citecolor=blue,
	linkcolor=blue,
	filecolor=blue,      
	urlcolor=blue,
}

\newcommand{\beq}{\begin{equation}}
\newcommand{\eeq}{\end{equation}}
\newcommand{\eml}{\end{mathletters}}

\newcommand{\be}{\begin{equation}}
\newcommand{\ee}{\end{equation}}
\newcommand{\bea}{\begin{eqnarray}}
\newcommand{\eea}{\end{eqnarray}}
\newcommand{\nn}{\nonumber\\}
\newcommand{\oh}{\frac{1}{2}}
\newcommand{\of}{\frac{1}{4}}

\newcommand{\ov}{\overline}

\begin{document}
	\title{Semiclassical propagator approach for emission processes.
	\\ I. Two-body non-relativistic case}

\author{S.A. Ghinescu $^{1,2}$}
\email[Corresponding author:]{stefan.ghinescu@nipne.ro}

\author{D.S. Delion $^{1,2,3,4}$}

\affiliation{
$^1$ "Horia Hulubei" National Institute of Physics and Nuclear Engineering, \\
30 Reactorului, POB MG-6, RO-077125, Bucharest-M\u agurele, Rom\^ania \\
$^2$ Department of Physics, University of Bucharest,
405 Atomi\c stilor, POB MG-11, RO-077125, Bucharest-M\u agurele, Rom\^ania\\
$^3$ Academy of Romanian Scientists, 3 Ilfov RO-050044,
Bucharest, Rom\^ania \\
$^4$ Bioterra University, 81 G\^arlei RO-013724, Bucharest, Rom\^ania}
\date{\today}

\begin{abstract}
We compare the coupled channels procedure to the semiclassical approach
to describe two-body emission processes, in particular $\alpha$-decay, from deformed nuclei
within the propagator method.
We express the scattering amplitudes in terms of a propagator matrix, 
describing the effect of the deformed field, multiplied by the ratio between internal 
wave function components and irregular Coulomb waves. In the spherical case
the propagator becomes diagonal and scattering amplitudes acquire the well-known form.
We describe a more rigorous formulation of the 3D semiclassical approach, corresponding to
deformed potentials, which leads to the exact results and we also compare 
them with the much simpler expressions given by the Angular Wentzel-Krames-Brillouin 
(AWKB) and Linearized WKB (LWKB) with its approximation, known as Fr\"oman WKB (FWKB) method.
We will show that LWKB approach is closer than AWKB to the exact coupled-channels formalism.
An analysis of alpha-emission from ground states of even-even nuclei evidences the
important role played by deformation upon the channel decay widths.
\end{abstract}

	\maketitle
	
	\setcounter{equation}{0}
	
	\section{Introduction}
\setcounter{equation}{0} 
\renewcommand{\theequation}{1.\arabic{equation}} 

The exact description of emission processes is provided by outgoing solutions in continuum
of the equation of motion. When the masses of the emitted particles are much larger than 
the energy release (Q-value) the non-relativistic Schr\"odinger equation is used, while in the other case
a relativistic approach is employed within the Klein-Gordon equation for boson emission and 
Dirac equation for fermion emission.
The case of non-relativistic two-body processes refers by the one proton emission,
alpha and heavy cluster decays $P\rightarrow D+C$, while two-proton emission $P\rightarrow D+p+p$
belongs to the field of non-relativistic three-body dynamics. 
The most important relatistic three-body emission processes is given by the 
$\beta^+$ decay $p\rightarrow n+e^+ + \nu$, where the positron mass has a comparable value 
to the Q-value but it penetrates a very large barrier, comparable to the proton emission case.
The exact solutions for deformed emitters in all these cases are provided within the coupled channels (CC)
approach with an outgoing asymptotics \cite{Del10}.
All these processes are basically described by the quantum penetration of a particle/cluster 
through an internal nuclear plus an external Coulomb barrier,
characterized by a relative small ratio between the Q-value and barrier height.
In this case semiclassical solutions provide very good approximations
and our purpose is to analyze such solutions in the most general cases,
by applying the so-called propagator method, already described for the CC approach in Ref. \cite{Del10}.

We will describe in this paper the two-body non-relativistic emission,
where a very good approximation is given by the semiclassical 
Wentzel-Kramers-Brillouin (WKB) approach \cite{Gam28,Con28,Del15}.
The problem of a formulating a general three-dimensional (3D) WKB theory for systems lacking spherical symmetry 
has a long history. The first successful attempt is due to Fr\"oman \cite{Fro57} who obtained 
a "semi-analytic" expression for the wave-function of an alpha particle inside a large barrier 
using geometrical considerations. His attempt is not, however, free from caveats especially due to the \textit{intuitive}
approach he followed. We present here a more rigorous formulation which leads to the
so called Linearized WKB (LWKB), which has as a particular case the Fr\"oman method. 
We also compare them with the much simpler expression which has seen extensive use by many authors 
\cite{Del15,Ste96}.
We will refer to this method as "Angular WKB" or AWKB, in short. 
In the end we show that both methods agree with  the exact coupled-channels formalism for small to reasonable deformations.
We will apply these considerations in the case of alpha decays to ground and excited states.
	
\section{Mathematical formulation}
\setcounter{equation}{0} 
\renewcommand{\theequation}{2.\arabic{equation}} 

	To give a full account of all steps we begin with the spherically symmetric problem,
	the reason being that the centrifugal term in the potential appears naturally when
	one builds up the deformed solution as an extension of the spherical one.

\subsection{Spherical emitters}
Let us consider a binary emission process
\bea
\label{decay}
P(J_i)\rightarrow D(J_f)+C(L)~
\eea
where $J_{i/f}$ denotes the initial/final ${\rm spin^{parity}}$ of the parent (P)/daughter (D) nucleus and $L$ the
angular momentum carried by the emitted cluster (C). For simplicity we consider the cluster with a boson
structure (an alpha particle or heavier cluster). We will also assume an initial ground state $J_i=0$,
leading to $J_f=L$, i.e. a coupled daughter-cluster dynamics with the total spin $L\otimes L=0$.
The Schr\"odinger equation governs the dynamics of the binary D+C system
 inside a spherically symmetric potential barrier $V_0(r)$
	\begin{equation}
	\label{eq:Schodinger_Initial}
	\left[-\frac{\hbar^2}{2\mu}\Delta + V_0(r)\right]\Psi_0(\textbf{r}) = E\Psi_0(\textbf{r})~,
	\end{equation}
	where $\textbf{r}=(r,\theta,\phi)$ denotes
	the position vector of the cluster in the center of mass (CM) of the system in spherical coordinates 
	and $\mu = m_Cm_D/(m_C+m_D)$ defines the daughter-cluster reduced mass.
	The generalisation to the emission of fermions is straightforward.
Notice that inside the external Coulomb barrier the standard multipole expansion 
\bea
\label{expans}
\Psi_0({\bf r})=\sum_L\frac{f_L(r)}{r}Y_{L0}(\theta)~,
\eea
leads at a large distance to the following equations for radial components
\bea
\label{asymp}
\left[-\frac{d^2}{d\rho^2}+\frac{L(L+1)}{\rho^2}+\frac{\chi}{\rho}-1\right]f_l(r)=0~,
\eea
depending upon the Coulomb parameter
\bea
\label{Coul}
\chi=\frac{2Z_DZ_C}{\hbar v}
~,
\eea
and reduced radius
\bea
\label{rho}
\rho&=&k r,~~~~
k=\sqrt{\frac{2\mu E}{\hbar^2}}~.
\eea
	We employ the semiclassical ansatz by writing
	\begin{equation}
	\Psi_0(\textbf{r}) \equiv \exp\left[\frac{i}{\hbar}S_0(\textbf{r})\right]~.
	\end{equation}
	Upon inserting this expression in Eq.~(\ref{eq:Schodinger_Initial}), we obtain
	\begin{equation}
	\label{eq:Schrodinger_for_expo}
	-\frac{i\hbar}{2\mu}\Delta S_0(\textbf{r}) + \frac{1}{2\mu}\left[\nabla S_0(\textbf{r})\right]^2 
	+ V_0(r) = E~,
	\end{equation}
	where $\Delta$ and $\nabla$ denote the laplacian and gradient respectively in spherical coordinates.
	
	The semiclassical prescription requires the exponent $ S_0(\textbf{r}) $ to be expanded 
	in powers of $ \hbar $ as $ S_0(\textbf{r}) = S_0^{(0)}(\textbf{r})  + \hbar S_0^{(1)} (\textbf{r})$. 
	We plug the expansion in Eq.~(\ref{eq:Schrodinger_for_expo}) and group coefficients of 
	equal powers of $ \hbar $ to obtain the following system of equations
	\begin{align}
	\label{eq:WKB_system}
	\begin{aligned}
	\hbar^0:& \left(\nabla S_0^{(0)}(\textbf{r})\right)^2 = -K_0^2(r)\\
	\hbar^1:& -\frac{i}{2}\Delta S_0^{(0)}(\textbf{r}) + (\nabla S_0^{(0)}(\textbf{r}))(\nabla S_0^{(1)}(\textbf{r})) = 0~,
	\end{aligned}
	\end{align}
	where we defined the "radial dependent momentum"
	\begin{equation}
		\label{eq:K0_definition}
		K_0(r)\equiv\sqrt{2\mu E\left[\frac{V_0(r)}{E}-1\right]}~.
	\end{equation}
	We show in the Appendix that an "outgoing" solution of this system is given by
\bea
\label{Psi0}
\Psi_0(r,\theta) &=& \sum_{L}
\frac{c_LY^{(\textrm{WKB})}_{L0}(\theta)}{\sqrt{K_{0,L}(r)}}\exp\left[\int_{r}^{r_{2}}drK_{0,L}(r)\right]~,
\nn
\eea
	where $Y^{(WKB)}_{L0}$ is the spherical harmonic $Y_{L0}$ in the WKB approximation, $c_L$
	are some constants, $r_{2}$ is the external turning point defined as the largest solution 
	of the equation $V_0(r_{2})=E$ and 
	\begin{align}
		\begin{aligned}
		K_{0,L}(r) &\equiv \sqrt{2\mu E\left[\frac{V_0(r)}{E} - 1 + 
		\frac{\left(L+\frac{1}{2}\right)^2}{k^2r^2}\right]}~.
		\end{aligned}
	\end{align}
	Note that we have dropped the $\phi$ dependence, which appears only in the form of a phase 
	since the potential is spherically symmetric. This simplifies expressions in both the spherical
	and deformed cases without loss of generality.

\subsection{Deformed emitters}
	
We turn now to solving the deformed problem. In the laboratory system of coordinates
the dynamics of the emission process (\ref{decay}) is described by the following Schr\"odinger equation
\bea
\left[\widehat{\bf H}({\bf R})+\widehat{\bf H}_D(\Omega)+V({\bf R},\Omega)\right]\Phi({\bf R},\Omega)
=E\Phi({\bf R},\Omega)~,
\nn
\eea
where $\widehat{\bf H}({\bf R})$ denotes the Hamiltonian of the daughter-cluster motion 
depending on the relative coordinate ${\bf R}=(r,\widehat{R})$ and
$\widehat{\bf H}_D(\Omega)$ describes the internal daughter motion depending on its coordinate $\Omega$,
which is given by Euler angles for rotational motion.
We will consider an axially symmetric daughter-cluster interaction
which can be estimated within the double folding procedure \cite{Ber77,Sat79,Car92} by the following expansion
\bea
\label{poten}
V({\bf R},\Omega)&=&V_0(r)+\sum_{\lambda>0}V_{\lambda}(r)\sqrt{\frac{4\pi}{2\lambda+1}}
\left[Y_{\lambda}(\Omega)\otimes Y_{\lambda}(\widehat{R})\right]_0
\nn&=&
V_0(r)+\sum_{\lambda>0}V_{\lambda}(r)Y_{\lambda 0}(\widehat{r})
\nn&\equiv&
V_{0}(r) + V_{d}({\bf r})~,
\eea
where $\widehat{r}$ is the daughter-particle angle, defining the intrinsic system of coordinates ${\bf r}=(r,\widehat{r})$,
with $V_0$, the isotropic component (monopole), and $V_{d}({\bf r})$, the purely anisotropic part.
We expand solution in the intrisic system to obtain in a standard way the coupled system of equations. 
By neglecting the off-diagonal Coriolis terms within the so-called adiabatic approach
one obtains at large distances a similar to (\ref{asymp}) form, 
but with different Coulomb parameters and reduced radii in each channel \cite{Del10,Fro57}
\bea
\label{energ}
\chi_L&=&\frac{\chi}{\epsilon_L}
\nn
\rho_L&=&\rho\epsilon_L
\nn
\epsilon_L&\equiv&\sqrt{1-\frac{E_L}{E}}~,
\eea 
where $E_L$ denotes the excitation energy of the daughter nucleus. This corresponds to
the energy replacements $E\rightarrow E-E_L$ in each channel.

In order to analyze the specific features of the deformed WKB approach 
we will first neglect the excitations energies of the daughter nucleus,
which will be considered later in applications.
	The corresponding Schrodinger equation now reads
	\begin{equation}
		\label{eq:Sch_deformed}
		\left[-\frac{\hbar^2}{2\mu}\Delta + V_0(r)+V_d(\textbf{r})\right]\Psi(\textbf{r}) = E\Psi(\textbf{r})~.
	\end{equation}
	We propose a semiclassical ansatz similar to the one in the spherical case
	\begin{equation}
		\label{eq:WKB_deformed_ansatz}
		\Psi(\textbf{r})=\exp\left[\frac{i}{\hbar}S(\textbf{r})\right]~,
	\end{equation}
	from which we obtain the deformed equivalent of the system in Eq.~(\ref{eq:WKB_system}) by making 
	again the expansion in powers of $\hbar$ as $S(\textbf{r}) = S^{(0)}(\textbf{r}) + \hbar S^{(1)}(\textbf{r})$
	\begin{align}
		\label{eq:WKB_system_deformed}
		\begin{aligned}
		\hbar^0:& \left(\nabla S^{(0)}(\textbf{r})\right)^2 = -K^2(\textbf{r})\\
		\hbar^1:& -\frac{i}{2}\Delta S^{(0)}(\textbf{r}) + (\nabla S^{(0)}(\textbf{r}))(\nabla S^{(1)}(\textbf{r})) = 0~,
		\end{aligned}
	\end{align}
	where we have defined
	\begin{equation}
		\label{eq:DefK_definition}
		K(\textbf{r}) \equiv \sqrt{2\mu E \left[\frac{V_0(r)}{E}-1+\frac{V_{d}({\bf r})}{E}\right]}~.
	\end{equation}
	The approach followed by Fr\"oman to solve Eqs.~(\ref{eq:WKB_system_deformed}) is known today as the \textit{linearization} of the Eikonal equation which applies to $S^{(0)} $ in our case. This approximation consists
in isolating the spherical part $K_0(r)$ defined by Eq.~(\ref{eq:K0_definition}) in the first equation ~(\ref{eq:WKB_system_deformed}). We call this approach as Linearized WKB (LWKB).
This can be achieved through the binomial approximation if $ V_d $ is small compared with $ V_0 $ (in the following we omit the spatial variables trusting no ambiguity arises)
	\begin{align}
		\label{eq:KWKB_binomial}
		\begin{aligned}
		K({\bf r}) &= \sqrt{2\mu E\left(\frac{V_0}{E}-1\right)}\sqrt{1+\frac{V_d/E}{V_0/E-1}}\\
		&\approx \sqrt{2\mu E\left(\frac{V_0}{E}-1\right)}\left(1 + \frac{1}{2}\frac{V_d/E}{V_0/E-1}\right)\\
		&\equiv K_0 + \frac{\Delta K}{K_0}~,
		\end{aligned}
	\end{align}
	where we have defined
	\begin{equation}
		\label{eq:deltaKWKB_def}
		\Delta K \equiv \frac{1}{2}2\mu V_d~.
	\end{equation}
As we mentioned, Fr\"oman WKB approach (FWKB) is a particular case of LWKB and it corresponds 
to a pure Coulomb potential of a deformed nucleus with a sharp density distribution. 
In this case various multipoles of $V_d$ have closed analytic expressions. 
	
	It is clear now that, since we have isolated the spherical contribution, we can use the solution from its associated problem. We write $S^{(0)}(\textbf{r})$ as
	\begin{equation}
		\label{eq:S0_deformed_definition}
		S^{(0)}(\textbf{r}) = S^{(0)}_{0}(\textbf{r}) + D(\textbf{r})~,
	\end{equation}
	where $S^{(0)}_0$ is the solution of the spherical problem given by Eq. (\ref{Psi0}), and $D(\textbf{r})$ 
	is the correction arising from the potential deformation. Then we replace this definition together with
	 Eqs.~(\ref{eq:KWKB_binomial},\ref{eq:deltaKWKB_def}) inside Eq.~(\ref{eq:WKB_system_deformed}) and obtain
	\begin{align}
	\begin{aligned}
		\left(\nabla S^{(0)}_{0}\right)^2 + \left(\nabla D\right)^2 + 2\left(\nabla S^{(0)}_{0}\right)
		\left(\nabla D\right) =\\
		 -\left(K_0^2 + \frac{\Delta K^2}{K_{0}^2} + 2 \Delta K\right)~.
	\end{aligned}
	\end{align}
	
	The essence of the linearized eikonal approximation consists in 
	neglecting terms of powers higher than $ 1 $ in both $ \Delta K $ 
	and $ \nabla D $. We enforce now this idea and, after small 
	simplifications, we obtain 
	\begin{equation}
		\label{eq:WKB_eqD0}
		(\nabla S^{(0)}_{0})(\nabla D) = -\Delta K~.
	\end{equation}
	
	As shown in Appendix, the partial derivatives of $ S^{(0)}_{0} $ 
	are given by
	\begin{align}
		\label{eq:S0_derivatives}
		\begin{aligned}
		\frac{\partial S^{(0)}_{0}}{\partial \theta} &= \left(L+\frac{1}{2}\right)\hbar \\
		\frac{\partial S^{(0)}_{0}}{\partial r} &= \pm i K_{0,L}(r)~,
		\end{aligned}
	\end{align}
	where $ L $ is the angular momentum quantum number.
	
	We see now that our problem reduces to solving the equation
	\begin{equation}
		\label{eq:FromanEq_exponent}
		iK_{0,L}(r) \frac{\partial D}{\partial r} + \frac{\left(L+\frac{1}{2}\right)\hbar}{r^2}
		\frac{\partial D}{\partial \theta} = -\Delta K~.
	\end{equation}

\begin{figure}[ht] 
\begin{center} 
\includegraphics[width=9cm]{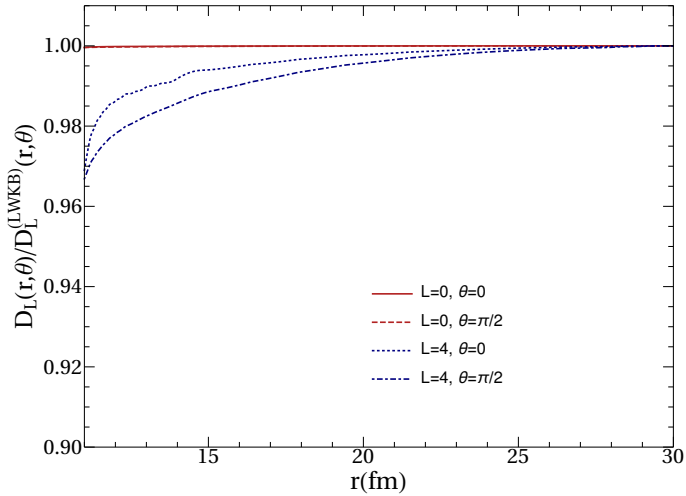} 
\caption{
Ratio between the solution of eq. (\ref{eq:FromanEq_exponent}) and LWKB approximation
versus radius for $L=0,~4$ and $\theta=0,~\pi/2$.
}
\label{fig1}
\end{center} 
\end{figure}
	
	This equation does not have a closed form solution unless the deformed potential is of the form $ V_d(r,\theta) = \mathcal{V(\theta)}/r^2 $, which is not the case for axial deformations. 
Consequently, the approximation used is that even though the potential is no longer spherically symmetric, the classical trajectory of the emitted cluster would still be a straight line and one can integrate this system radially by setting formally $ \partial D/\partial \theta = 0$. This approximation is somewhat justified also by the coefficients of the two partial derivatives: far away from the turning points $K_{0,L}(r)$ is of the order 1, while $(L+1/2)/r \propto 0.1$. 
Indeed, we solved numerically equation Eq.~(\ref{eq:FromanEq_exponent}) and plotted the ratio between exact and LWKB solutions
in Fig. \ref{fig1}. One notices that the deviation with respect to the LWKB solution is very small.

In this case, the Fr\"oman correction on each channel can be integrated starting from far away from the nucleus, say from a point $r=r_0$, where the field is spherical (hence $ D = 0$ for all $ l $) and the exponent becomes
	\begin{equation}
		\label{eq:Solution_D0}
		\frac{i}{\hbar}D(r,\theta) =  - \frac{1}{\hbar}\int_{r_0}^{r}dr'
		 \frac{\Delta K(r',\theta)}{K_{0,L}(r')}~.
	\end{equation}
	
	We note here that this is the correct use of the WKB approximation since it gives the expected asymptotic behavior, while in \cite{Fro57} the author performs the integration starting from the nuclear surface towards the turning point. This observation is useful, however, only if one desires to compute the wave-function of the alpha particle inside the barrier at a specific point. By contrast, if we wish to compute only the penetrability, both expressions are equally valid.
	
	The last step would be to consider the effect of the deformation on the quantum term $ S^{(1)} $ in Eq.~(\ref{eq:WKB_system_deformed}), but this proves to be quite small compared to what we have discussed already so we omit the correction, keeping only the spherical part $S^{(1)}(\textbf{r})\approx S^{(1)}_{0}(\textbf{r})$. The procedure is the same, the derivative with respect to $ \theta $  is neglected in the $ S_1 $ term and the integration is carried out radially. This approximation performs rather well as we will show in the following section.
	
	We turn our focus on the AWKB method. In the first few paragraphs of this chapter we claimed it is more elegant than the one of Fr\"oman and now we will provide some arguments. In order not to repeat all the equations we refer the reader to the system from Eq.~(\ref{eq:WKB_system_deformed}). If we do not attempt to linearlize this equation, the only way towards a "semi-analytic" expression is again radial integration. We set $ \partial S^{(0)}/\partial \theta = 0 $, but this is not enough. We do not retrieve in this way the angular momenta enumeration, hence we still have to separate the spherical contribution. We can do this by writing
	\begin{equation}
		\label{eq:KAWKB_decomposed}
		K^2(\textbf{r}) = K_0^2(r) + \Delta K^{\mathrm{(AWKB)}}(\textbf{r})~,
	\end{equation}
	where we have defined
	\begin{equation}
		\label{eq:DeltaKAWKB_definition}
		\Delta K^{\mathrm{(AWKB)}}(\textbf{r}) = 2\mu V_{d}(\textbf{r})~.
	\end{equation}
	We now use the definition of $S^{(0)}$ from Eq.~(\ref{eq:S0_deformed_definition}) without neglecting any term to write
	\begin{equation}
		\label{eq:DWKB_diffeq}
		(\nabla S^{(0)}_{0})^2 + (\nabla D)^2 +2(\nabla S^{(0)}_{0})(\nabla D) = -K_0^2 - \Delta K^{\mathrm{(AWKB)}}~.
	\end{equation}
	As per Eq.~(\ref{eq:WKB_system}), $(\nabla S^{(0)}_{0})^2 = -K_0^2$, and if we set $\partial D/\partial\theta = 0$ again, we obtain
	\begin{equation}
		\left(\frac{\partial D}{\partial r}\right)^2 + 2\frac{\partial S_{0}^{(0)}}{dr}\frac{\partial D}{\partial r} = -\Delta K^{(\mathrm{AWKB})}~.
	\end{equation}
	The derivative of $S^{(0)}_{0}$ with respect to $r$ is given in Eq.~(\ref{eq:S0_derivatives}) and we solve this quadratic equation for $dD/dr$ as
	\begin{align}
		\label{eq:D_AWKB_diffeq}
		\begin{aligned}
		\frac{\partial D}{\partial r} &= \mp i K_{0,L} \pm i \sqrt{K_{0,L} + \Delta K^{(\mathrm{AWKB})}}\\ 
		&\equiv \mp iK_{0,L}(r) \pm i K_{L}(\textbf{r})~,
		\end{aligned}
	\end{align}
	where we have denoted
	\begin{equation}
		\label{eq:KL_deformed_definition}
		K_{L}(\textbf{r}) = \sqrt{2\mu E\left(\frac{V_0}{E}-1 + \frac{\left(L+\frac{1}{2}\right)^2}{k^2 r^2} + \frac{V_d(\textbf{r})}{E}\right)}~.
	\end{equation}
	Upon integrating the last equation, we retrieve the well-known (but not proved) inclusion of the centrifugal potential in the 3D WKB exponent
	\begin{equation}
		\label{eq:Solution_D0_explicit}
		\frac{i}{\hbar}D(r,\theta) = 
-\frac{i}{\hbar}S^{(0)}_0 (r) + \frac{1}{\hbar} \int_{r}^{r_{0}} dr' K_L(r',\theta)~,
	\end{equation}
	for some radius $r_{0} > r$ where the function is known. So it turns out that the "mixed" representation where the centrifugal term is included a priori is actually less approximate than Fr\"oman's method, at least in principle. Now, regarding the second term in the expansion, in this case it is given as an extension of the spherical case
	\begin{equation}
		S^{(1)}(r,\theta) = \frac{i}{2}\ln\sqrt{K_{L}(r,\theta)}~.
	\end{equation} 
	This result follows if one integrates radially the original system of equations with the centrifugal potential inserted in the exponent as described above.
	
\subsection{Propagator method}

	Now, since we have build the wave-functions at all coordinates $(r,\theta)$, we can compare the WKB results with the exact 
coupled channels (CC) one. To achieve this we must build the fundamental matrix of solutions in the WKB case.
The exact CC fundamental matrix of solutions is defined by the following asymptotics \cite{Del10}
\bea
\label{funmat}
{\cal H}^{(CC)}_{LL'}(r)\rightarrow_{r\rightarrow\infty} H_{L}^{(+)}(kr,\chi)\delta_{LL'}~,
\eea
in terms of the outgoing Coulomb-Hankel spherical waves 
$H_{L}^{(+)}(\chi,kr)=G_{L}(\chi,kr)+iF_{L}(\chi,kr)$.
Thus, each column of the fundamental matrix of solutions is obtained by integrating backwards
the coupled system of differential equations, starting with above mentioned asymptotic value.
Notice that inside the Coulomb barrier this matrix has practically real values, due to the
fact that here one has $G_{L}(\chi,kr)>>F_{L}(\chi,kr)$. Therefore in practical calculations
one uses only the irregular Coulomb wave at large distance.
The general solution with a given angular momentum is built as a superposition of columns
\bea
\label{fL}
f_L(r)&=&\sum_{L'}{\cal H}^{(CC)}_{LL'}(r)N_{L'}
\nn
&\rightarrow_{r\rightarrow\infty}& N_LH_{L}^{(+)}(kr,\chi)~.
\eea
This expression can be used to find scattering amplitudes $N_L$ in terms of components
of the internal function at some radius $r$ inside the barrier
by using the matching condition $f_L^{(int)}(r)=f_L(r)$
\bea
\label{NL}
N_L=\frac{1}{H^{(+)}_L}\sum_{L'}{\cal K}^{(CC)}_{LL'}(r)f^{(int)}_{L'}(r)~,
\eea
where we introduced the propagator matrix \cite{Del10} as follows
\bea
\label{KLL}
{\cal K}^{(CC)}_{LL'}(r)&\equiv& H^{(+)}_L(\chi,kr)\left[{\cal H}^{(CC)}_{LL'}(r)\right]^{-1}
\nn
&\approx& G_L(\chi,kr)\left[{\cal H}^{{(CC)}}_{LL'}(r)\right]^{-1}~,
\eea
with the following property
\bea
{\cal K}^{(CC)}_{LL'}(r)\rightarrow_{V_d\rightarrow 0}\delta_{LL'}~,
\eea
which takes place for a sperical interaction, or for a deformed interaction at large distance
where it becomes spherical.
 
We observe from Eqs.~(\ref{eq:Solution_D0}) and (\ref{eq:Solution_D0_explicit}) that the complete wave-function in both cases can be written as
	\begin{equation}
		\label{eq:WKB_full_wavefunction}
		\Psi(r,\theta) = \sum_{L}\psi_{L}(r,\theta)Y_{L}(\theta)~,
	\end{equation}
	where $\psi_{L}(r,\theta)$ are built up using the WKB functions
	\begin{equation}
		\label{eq:psi_wkb}
		\psi_{L}(r,\theta) = \exp\left\lbrace\frac{i}{\hbar}\left(S^{(0)}(r,\theta) + \hbar S^{(1)}(r,\theta)\right)\right\rbrace~,
	\end{equation}
	with both $S^{(0)}$ and $S^{(1)}$ depending on $L$, as we have shown. Then, the radial components of the complete wave-function are given by
	\begin{equation}
		\Psi_{L}(r)= \sum_{L'} \int d\Omega Y_{L} \psi_{L'}(r) Y_{L'}~,
	\end{equation}
	which can readily be translated to the fundamental matrix with the asymptotics (\ref{funmat}) as
	\begin{equation}
		\label{eq:fundam_matrix_wkb}
		\mathcal{H}(r)\equiv\mathcal{H}_{L,L'}(r) = \int d\Omega Y_{L}\psi_{L'}(r)Y_{L'}~.
	\end{equation}
	
	We now particularize Eq.~(\ref{eq:fundam_matrix_wkb}) in the AWKB and LWKB approaches. For the AWKB approximation, we have
	\begin{align}
		\begin{aligned}
		\label{eq:fundam_mat_Delion}
		\mathcal{H}^{(\textrm{AWKB})}_{L,L'}(r) &=\int d\Omega Y_{L}(\Omega) Y_{L'}(\Omega) \\
		&\times\exp\left\lbrace k\int_{r}^{r_{2,L}(\theta)}dr'K_{L'}(r',\theta)\right\rbrace~,
		\end{aligned}
	\end{align}
	where $r_{2,L}(\theta)$ are the angle dependent external turning points, i.e. the largest root of the equation
	\begin{equation}
		\label{eq:ang_tp}
		\frac{V_{0}(r)}{E} - 1 + \frac{V_{d}(r,\theta)}{E} + \frac{\left(L+\frac{1}{2}\right)^2}{k^2r^2} = 0 ~,
	\end{equation}
	at each angle. For LWKB approach, we can isolate the spherical contribution and write
	\begin{align}
		\begin{aligned}
		\label{eq:fundam_mat_Froman}
		&\mathcal{H}^{(\textrm{LWKB})}_{L,L'}(r) = G_{L'}(r)
		\\&\times\int d\Omega Y_{L}(\Omega) Y_{L'}(\Omega)
		\exp\left[\frac{i}{\hbar}D_{0,L}(r,\theta)\right]
		\end{aligned}
	\end{align}
where, according to Eq. (\ref{eq:Solution_D0})
\bea
\label{D0L}
D_{0,L}(r,\theta)&\equiv-\frac{k}{2}\int_{r}^{r_{2,L}}dr'\frac{\Delta K(r',\theta)}{K_{0,L}(r')}~,
\eea
is the deformed part of the exponential dependence generating the fundamental matrix of solutions.
	Here, $G_{L}(r)$ is the solution of the spherical problem and given by (see appendix)
	\begin{align}
		\label{eq:GL_defintion}
		\begin{aligned}
		G_{L}(r) &= \frac{1}{\sqrt{K_{0,L}(r)}}\exp\left [\int_{r}^{r_{2,L}} dr'K_{0,L}(r')\right]~,
		\end{aligned}
	\end{align}
	and $r_{2,L}$ is the \textit{spherical} external turning point, i.e. the largest solution of the equation
	\begin{equation}
		\label{eq:tp_spher}
		\frac{V_{0}(r)}{E}-1+\frac{\left(L+\frac{1}{2}\right)^2}{k^2r^2} = 0~.
	\end{equation}
	
	We note here that a somewhat similar treatment has been made by Stewart et al in \cite{Ste96} although the centrifugal potential is introduced \textit{ad hoc}, unlike in the AWKB approach of our paper. We also mention that in \cite{Fro57} the angular momentum dependence of the deformed correction is more approximate, while here we account for it completely. More precisely, the exponent in the deformed correction of Fr\"oman's original work contains the ratio $\Delta K/K_{0}(r)$, but our LWKB treatment gives the rigorous angular momentum dependence of the deformed term.
	
	A close inspection reveals that the LWKB deformed term in the fundamental matrix Eq.~(\ref{eq:fundam_mat_Froman}) can be regarded as a matrix which becomes unity in the case of 0 deformation. In order to compare the two approximations (with each other and with the 
CC equivalent), we have to force the spherical part in the AWKB method. This is done by defining 
	
	\begin{equation}
		\label{eq:Spherical_propag}
		\mathcal{G}_{L,L'}(r) \equiv G_{L}(r) \delta_{L,L'}~, 
	\end{equation}
	where $G_{L}$ are the solutions of the spherical problem (\ref{eq:Schodinger_Initial}). With this definition we can impose (in matrix form)
	\begin{equation}
		\label{eq:AWKB_propag_decomp}
		\mathcal{H}^{(\textrm{AWKB})} \equiv \mathcal{G} \Delta\mathcal{H}^{(\textrm{AWKB})}~,
	\end{equation}
	from which we get the deformed term $\Delta \mathcal{H}$ as
	\begin{align}
		\label{eq:AWKB_deformed_propag}
		\begin{aligned}
	 \Delta\mathcal{H}^{(\textrm{AWKB})}_{L,L'} 
		=\frac{1}{G_{L}}\mathcal{H}^{(\textrm{AWKB})}_{L,L'}~.
		\end{aligned}
	\end{align}
	A similar expansion can be performed for the CC fundamental matrix, but with $G_{L}^{(\textrm{CC})}$ and $\mathcal{H}^{(\textrm{CC})}$, the \textit{exact} spherical wave function for channel $L$ and the \textit{exact} deformed fundamental matrix respectively
	\begin{align}
	\label{eq:CC_deformed_propag}
	\begin{aligned}
	\Delta\mathcal{H}^{(\textrm{CC})}_{L,L'} 
	=\frac{1}{G^{(\textrm{CC})}_{L}}\mathcal{H}^{(\textrm{CC})}_{L,L'}~.
	\end{aligned}
	\end{align}
	
	For LWKB method, where the spherical term is already separated, we have 
	\begin{align}
		\label{eq:Froman_deformed_propag}
		\begin{aligned}
		 \Delta\mathcal{H}^{(\textrm{LWKB})}_{L,L'} =& \int d\Omega Y_{L}(\Omega) Y_{L'}(\Omega)
		 \exp\left[\frac{i}{\hbar}D_{0,L}(r,\theta)\right]~.
		\end{aligned}
	\end{align}
	
Notice that the propagator matrix (\ref{KLL}) in all cases is given by the obvious relation
\bea
\label{KHLL}
{\cal K}_{LL'}=\Delta {\cal H}^{-1}_{LL'}~.
\eea
	We could also perform a more symmetric decomposition of the AWKB and CC fundamental matrices by defining the matrix (we drop the AWKB and CC indexes in the reminder of this section)
	\begin{align}
		\label{eq:SpherMat_sym_decomp}
		\overline{\mathcal{G}}_{L,L'}(r) = \begin{pmatrix}
		\sqrt{G_{0}(r)} & 0 & 0 &...\\
		0 & \sqrt{G_{2}(r)} & 0 & ...\\
		... & ... & ... & ...
		\end{pmatrix}~,
	\end{align}
	with which the deformed fundamental matrices can be written as
	\begin{equation}
		\label{eq:DeformedPropag_symmetric}
		\mathcal{H} = \overline{\mathcal{G}} \Delta \overline{\mathcal{H}} ~\overline{\mathcal{G}}~.
	\end{equation}
	We invert the above equation and perform the sums in the matrix multiplication to obtain the analogs of  Eqs.~(\ref{eq:AWKB_deformed_propag},\ref{eq:CC_deformed_propag}) in the form
	\begin{equation}
		\label{eq:DeformedComponent_symmetric}
		\Delta \overline{\mathcal{H}}_{L,L'} = \frac{1}{\sqrt{G_{L}G_{L'}}}\mathcal{H}_{L,L'}~.
	\end{equation}

	\section{Numerical results}
\setcounter{equation}{0} 
\renewcommand{\theequation}{3.\arabic{equation}} 

In this section we compare the two approximations with the exact solution given by the CC method.
We also perform a systematic analysis of alpha decays from even-even emitters within
the deformed WKB approach.
	
\subsection{Coupled channels approach versus WKB}

\begin{figure}[ht] 
\begin{center} 
\includegraphics[width=9cm]{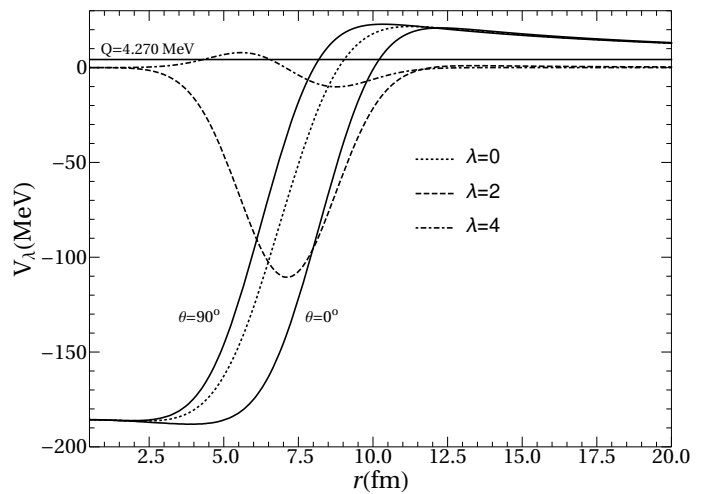} 
\caption{
Realistic alpha-daughter double-folding potential versus radius for the 
system $^{234}Th+\alpha$, plotted by a solid line for $\theta=0^o$
and $\theta=90^o$.
The multipoles are given by dotted ($\lambda$=0), dashed ($\lambda$=2) and
dot-dashed lines ($\lambda$=4).
The horizontal line corresponds to the $Q$-value of the emission process 
$^{238}U\rightarrow~^{234}Th+\alpha$ Q=E=4.270 MeV.
}
\label{fig2}
\end{center} 
\end{figure}

The realistic cluster-core interaction, given by the double-folding procedure \cite{Ber77,Sat79,Car92}, 
is plotted in Fig. \ref{fig2} versus radius for the binary deformed system $^{234}Th+\alpha$ 
with a quadrupole deformation $\beta_2$=0.215.
The two solid curves correspond to $\theta=0^o$ and $\theta=90^o$, respectivelly.
The multipoles in Eq. (\ref{poten}) are given by dotted ($\lambda$=0), dashed ($\lambda$=2) and
dot-dashed lines ($\lambda$=4).

First we compare the amplitudes at the matching radius following the recipe in \cite{Ste96} 
for the $^{238}U$ nucleus with scattering amplitudes (normalized to unity) $\mathcal{N} = \lbrace \sqrt{0.74}, \sqrt{0.25},
\sqrt{0.0004}\rbrace$.
By using the expression of the total decay width \cite{Del10}
\bea
\label{Gamma}
\Gamma&=&\sum_{L=even}\Gamma_L=\hbar v\sum_{L=even} \vert N_L\vert^2~,
\eea 
and (\ref{fL}) one can estimate the wave-function amplitudes at any point $r$
\bea
\label{fL}
f_L(r)&=&\sum_{L'}{\cal H}_{LL'}(r)\sqrt{\frac{\Gamma_{L'}}{\hbar v}}~.
\eea

	and must be normalized to unity as
\bea
f_{L}(r)\to \frac{f_{L}(r)}{\sum_{L'}|f_{L'}(r)|^2}~.
\eea
	We present the resulting amplitudes at $R_{m} = 12.1~\mathrm{fm}$ in Table~\ref{tab:wkb_amps_matching_radius} showing remarkably close results considering that Stewart et al computed these amplitudes at the internal turning point.
	
	\begin{table}[!htb]
		\caption{\label{tab:wkb_amps_matching_radius}$R_{\mathrm{m}} = 12.1\mathrm{fm}$}
		\begin{ruledtabular}
			\begin{tabular}{llll}
				$L$ & Stewart {\it et al} & LWKB & AWKB\\
				\hline\\
				$0$ & $+0.83$ & $+0.83$ & $+0.85$\\
				$2$ & $-0.55$ & $-0.54$ & $-0.52$\\
				$4$ & $-0.08$ & $-0.15$ & $-0.08$\\
			\end{tabular}
		\end{ruledtabular}
	\end{table}
\begin{figure}[ht] 
\begin{center} 

\includegraphics[width=8cm]{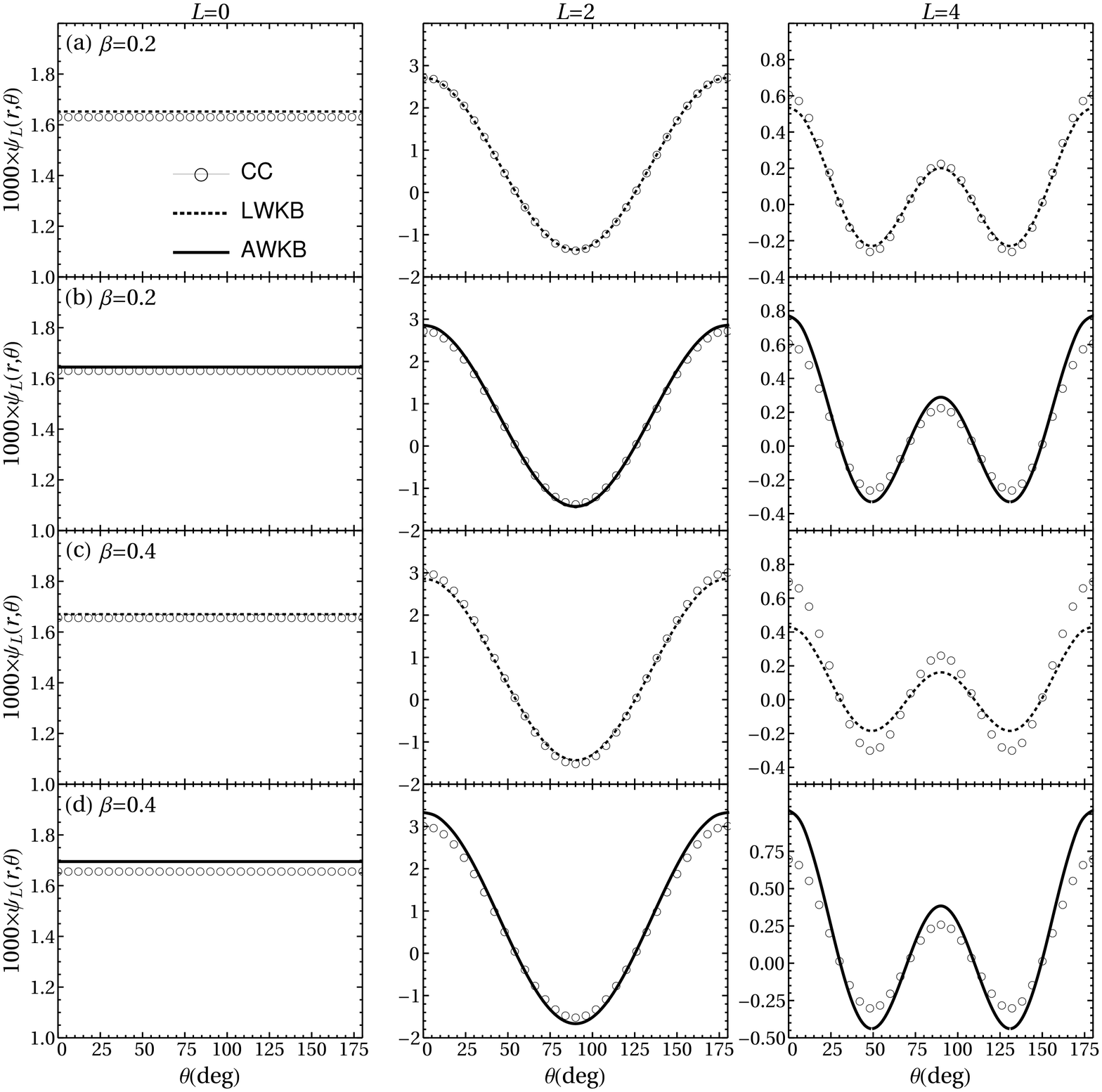} 
\caption{
The $L=0,~2,~4$ wave function components for CC (open symbols), LWKB (dots) and AWKB (solid line)
for $\beta_2=0.2$ (a) and (b) and $\beta_2=0.4$ (c) and (d), estimated by using the fundamental
matrix ${\cal H}$.
}
\label{fig3}
\end{center} 
\end{figure}

We have analyzed the accuracy of LWKB and AWKB approximations with respect to CC values.
The results are given in Fig \ref{fig3}, where we plotted the CC wave function components
for $L=0,~2,~4$ with open symbols for $\beta_2=0.2$ (a), (b) and $\beta_2=0.4$ (c), (d)
at the barrier radius. By dots we plotted LWKB components and by solid lines AWKB components.
 
\begin{figure}[ht] 
\begin{center} 
\includegraphics[width=8cm]{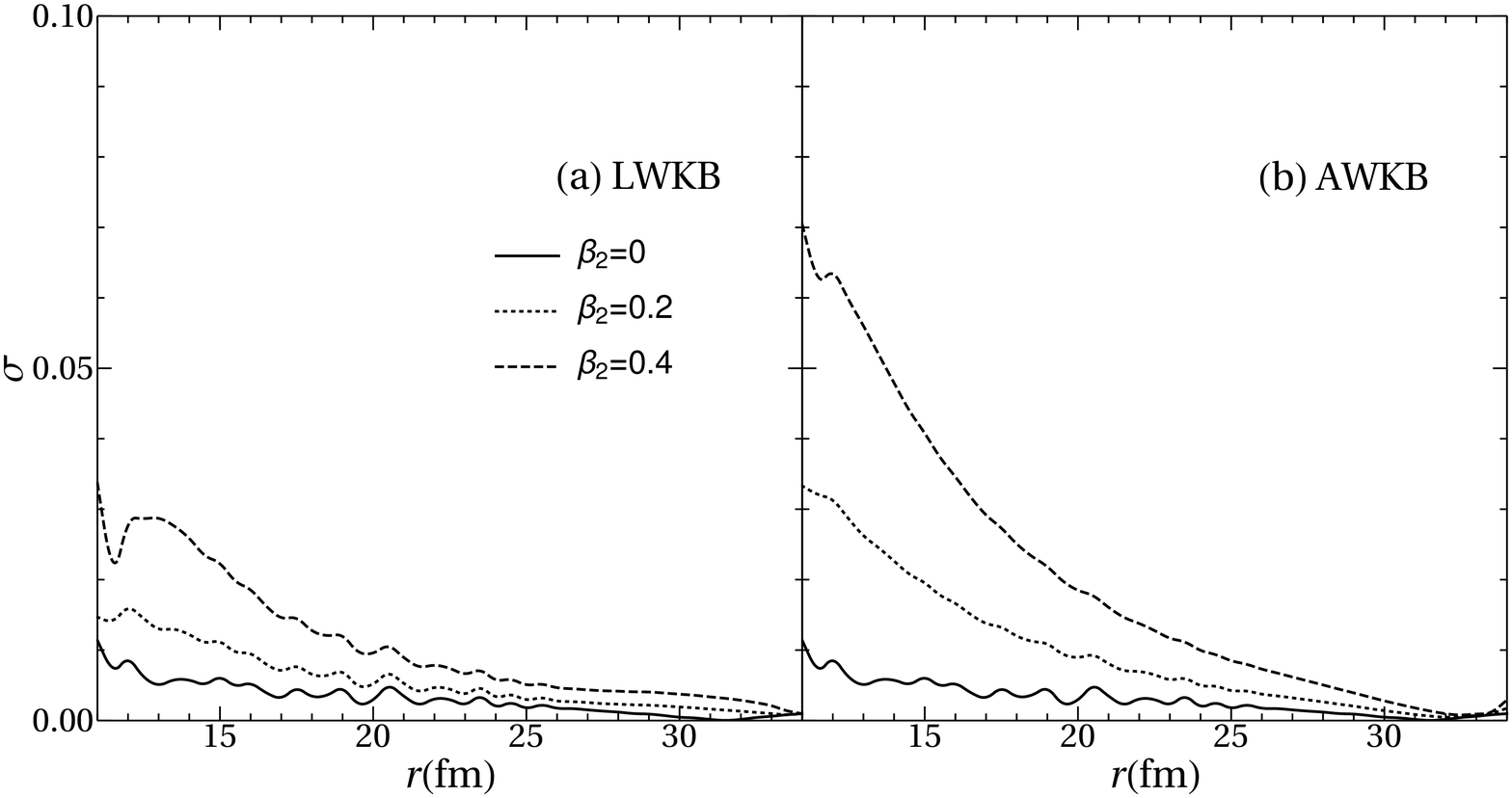} 
\caption{
(a) The relative error versus radius of the LWKB fundamental system of solutions
versus its coupled channels countert.
(b) Same as in (a) but for AWKB approach.
}
\label{fig4}
\end{center} 
\end{figure}

The results concerning the overal accuracy are shown in the Fig. \ref{fig4}.
As $\beta_2 $ increases, we see that LWKB approximation gives a reasonable relative error 
\bea
\sigma_X(r)&=&\sqrt{\frac{\sum_{LL'}\left[{\cal H}^{(X)}_{LL'}(r)-{\cal H}^{(CC)}_{LL'}(r)\right]^2}
{\sum_{LL'}\left[{\cal H}^{(CC)}_{LL'}(r)\right]^2}}~,
\nn
X&=&LWKB~,~~~AWKB~,
\eea
about $\sigma_{LWKB}(r_B)\sim 3\%$ for $\beta_2=0.4$ at the barrier radius, 
while AWKB corresponds to a twice larger, but still relative small, value $\sigma_{AWKB}(r_B)\sim 6\%$.

\subsection{Approximated interaction potential}

The region between the internal turning point and the barrier maximum can be approximated 
with a good accuracy by an inverted parabola
\bea
\label{eq:InternalPot_def}
V(r)-E&=&(V_B-E)\left[1-\left(\frac{r_B-r}{r_B-r_{1}}\right)^2\right]
\nn
&\equiv& V_{frag}(1-x^2)~,
\eea
in terms of the fragmentation potential
\bea
\label{Vfrag}
V_{frag}&=&V_B-E~,
\eea
and dimensionless coordinate
\bea
\label{x}
x&=&\frac{r_B-r}{r_B-r_{1}}~,
\eea
where $r_1$ denotes the internal turning radius.
The harmonic oscillator (ho) frequency parameter of the inverted parabola is given by
\bea
\label{hom}
\hbar\omega=\frac{1}{r_B-r_1}\sqrt{\frac{V_{frag}}{d_{\alpha}}}~,
\eea
in terms of the kinetic alpha-particle parameter
\bea
d_{\alpha}=\frac{2\mu_{\alpha}}{\hbar^2}\approx 0.192~MeV^{-1}fm^{-2}~.
\eea
Our previous analysis has shown that $\hbar\omega\sim$ 9 MeV \cite{Del20}.
The internal potential is matched to the external Coulomb potential $V_C(r)=2Z_D/r$
at the barrier maximum $r_B$, because the difference with respect to the exact
value is very small $V_B=0.94~V_C(r_B)$.

The scattering amplitude is given at the barrier radius by using (\ref{NL})
\bea
\label{scater}
N_L=\frac{1}{G_L(\chi,\rho_B)}\sum_{L'}{\cal K}_{LL'}(r_B)f^{(int)}_{L'}(r_B)~,
\eea
where the WKB estimate of the internal wave-function is given at 
the barrier radius $r_B$ by the Heel-Wheller ansatz \cite{Del20}
\bea
\label{fint}
f_{L}^{(int)}(V_{frag})&=&\frac{\sqrt{p_{L}}}{C_{N,L}}
\left(\frac{E}{V_{frag}}\right)^{\of}\exp\left(-S^{(0)}_N\right)~,
\nn
\eea
in terms of the spherical nuclear action
\bea
S^{(0)}_N&=&\frac{\pi V_{frag}}{2\hbar\omega}~,
\eea
and nuclear centrifugal term, given by the binomial approximation as follows
\bea
\label{CNL}
C_{N,L}&=&\exp\left[\left(L+\oh\right)^2d\right]
\nn
d&=&\frac{\delta}{2k}\sqrt{\frac{E}{V_{frag}}}
\left[ \frac{\delta}{\Delta^{2}r_{B}}+\frac{r_{B}}{\Delta^{3}} 
\left( \arctan{\frac{\delta}{\Delta}}-\frac{\pi}{2} \right) \right]
\nn
\delta&=&r_{B}-r_{1}~,~~~
\Delta=\sqrt{r_{B}^{2}-\delta^{2}}~.
\eea
The factor $p_L$ is called alpha-formation probability, which can be determined by experimental
channel widths.

Let us point out that we can use the potential, defined by (\ref{eq:InternalPot_def}), not only for the spherical part, 
but also for a deformed potential with $V_B=V_{B}(\theta)$ being the maximum barrier height
along the angle $\theta$, $r_B=r_{B}(\theta)$ its position and $r_{1}=r_{1}(\theta)$ 
the internal turning radius, which linearly depend upon the quadrupole deformation
\begin{equation*}
r_a(\theta) = r_{a,0}\left[1+b_a\beta_2 Y_{2,0}(\theta)\right]~,~~~a=B,~1~.
\end{equation*}
By using this ansatz we can easily estimate the deformed part if the internal action $D_N(\theta)$
defined by Eq. (\ref{D0L}).

The WKB estimate of the Coulomb spherical multipole in (\ref{scater}) is given by
\bea
G_L(\chi,\rho)=C_{C,L}\left(\cot\alpha\right)^{\oh}\exp\left(S_C^{(0)}\right)~,
\nn
\eea
in terms of the spherical Coulomb action
\bea
S_C^{(0)}&=&\chi\left(\alpha-\oh\sin~2\alpha\right)~,
\eea
and Coulomb angular momentum term, given in a standard way by the binomial approximation
\bea
\label{CCL}
C_{C,L}&=&
\exp\left[\left(L+\oh\right)^2c\right]
\nn
c&=&\frac{\tan\alpha}{\chi}~.
\eea
Here, we introduced the following parameter
\bea
\label{cosa}
\cos^2\alpha&=&\frac{\rho}{\chi}=\frac{E}{V_C(r)}~.
\eea
Notice that the above semiclassical estimate, valid for a pure Coulomb potential,
gives 3\% accuracy with respect to the exact function around the barrier region.

In order to estimate the fundamental and propagator matrix we used the separable LWKB 
approach (\ref{eq:Froman_deformed_propag}).
A simplified form is given by the Fr\"oman approach (FWKB) \cite{Fro57}, which neglects 
the centrifugal barrier in (\ref{D0L}) and uses a sharp density distribution at the nuclear surface 
$R=R_0\left[1+\beta_2Y_{20}(\theta)\right]$.
The result is proportional to the quadrupole deformation parameter and Legendre polinomial
\bea
\label{DC}
&&\frac{i}{\hbar}D^{(FWKB)}_C(\theta)=-\beta_2B(\chi,\rho)P_2(\cos\theta)
\nn
&&B(\chi,\rho)\equiv
\frac{\chi}{\sqrt{20\pi}}\sin 2\alpha\left(1+\sin^2\alpha\right)~.
\eea
Thus, the deformed part of the fundamental matrix (\ref{eq:Froman_deformed_propag}) 
within Fr\"oman approach is given by
\bea
\label{HLL}
\Delta{\cal H}^{(FWKB)}_{LL'}(\beta_2,\chi,\rho)&=&\int_{-1}^1d\cos\theta~\ov{P}_L(\cos\theta)\ov{P}_{L'}(\cos\theta)
\nn&\times&
\exp\left[-\beta_2B(\chi,\rho)P_2(\cos\theta)\right]~,
\nn
\eea
in terms of normalized Legendre polinomials 
\bea
\ov{P}_L(\cos\theta)=\sqrt{\frac{2}{2L+1}}P_L(\cos\theta)~.
\eea
Therefore the Fr\"oman propagator matrix (\ref{KHLL}) is given by
\bea
\label{KLL-Froman}
{\cal K}_{LL'}(\beta_2,\chi,\rho)&=&\left[\Delta{\cal H}^{(FWKB)}_{LL'}(\beta_2,\chi,\rho)\right]^{-1}
\nn
&=&\Delta{\cal H}^{(FWKB)}_{LL'}(-\beta_2,\chi,\rho)~.
\eea

\begin{figure}[ht] 
\begin{center} 
\includegraphics[width=8cm]{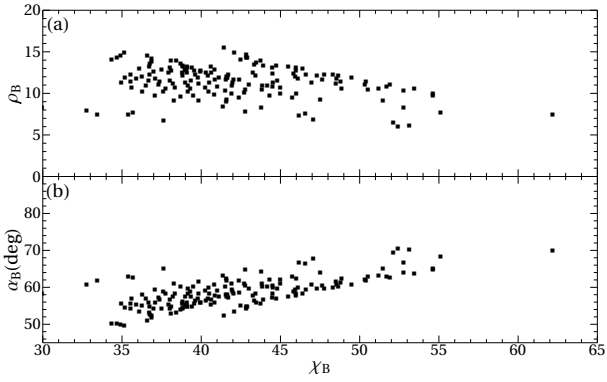} 
\caption{
(a) Reduced radius $\rho_B$ as a function the Coulomb parameter at the barrier radius
for alpha-transitions between ground states on even-even nuclei \cite{Del20}.
(b) Same as in (a), but for the angle $\alpha_B$.
}
\label{fig5}
\end{center} 
\end{figure}

\begin{figure}[ht] 
\begin{center} 
\includegraphics[width=8cm]{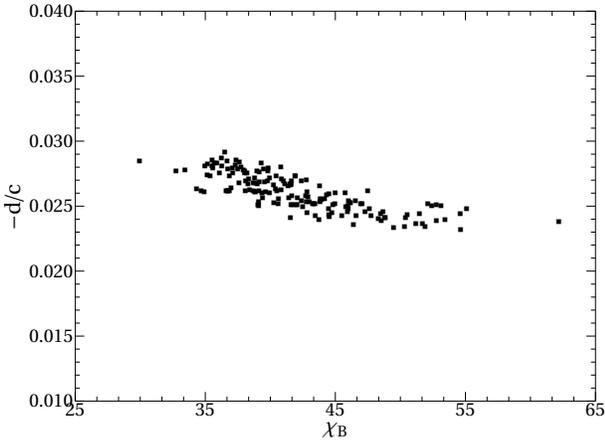} 
\caption{
Ratio between centrifugal temrs $-d/c$ versus Coulomb parameter.
}
\label{fig6}
\end{center} 
\end{figure}

\begin{figure}[ht] 
\begin{center} 
\includegraphics[width=8cm]{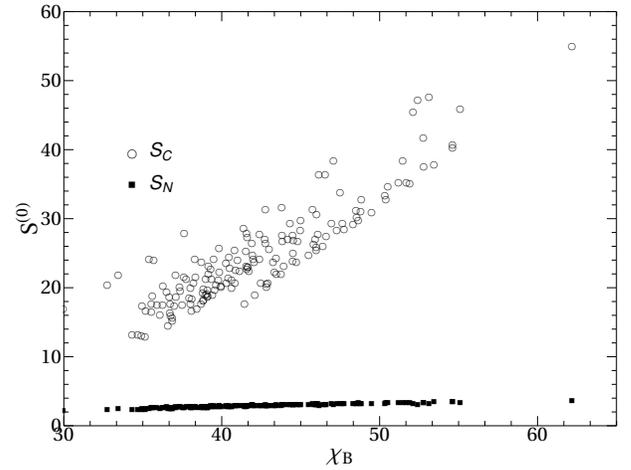} 
\caption{
Coulomb action $S_C^{(0)}$ (open symbols) nuclear action $S_N^{(0)}$ (dark symbols)
versus Colomb parameter for alpha-transitions between ground states of even-even nuclei.
}
\label{fig7}
\end{center} 
\end{figure}
	
\subsection{Alpha decay systematics}

We analyzed available experimental decay widths concerning alpha-transitions from 168
the ground state of even-even emitters with $J_i=0$ to final states with $J_f=L=0,~2,~4,...$. 
In Fig. \ref{fig5} (a) are given the values of the reduced radius
versus the Coulomb parameter  at the barrier radius $r_B=1.429(A^{1/3}+4^{1/3})$
by using transitions between ground states \cite{Del20}. 
In the panel (b) we plotted the corresponding angle $\alpha_B$ defined by Eq. (\ref{cosa}). 
They span the following intervals (except one isolated point)
\bea
\chi_B&\in&[33,~55]
\nn
\rho_B&\in&[5,~15]
\nn
\alpha_B&\in&[50^o,~70^o]~.
\eea
The last interval corresponds to a ratio between $Q$-value and the height of the Coulomb barrier
\bea
\frac{E}{V_B}\in[0.1,~0.4]~,
\eea
proving that the WKB approximation is very good for this kind of emission processes.

We analyzed the contribution of nuclear (\ref{CNL}) and Coulomb centrifugal factors (\ref{CCL}).
>From Fig. \ref{fig6} we notice that the nuclear term is much smaller that its Coulomb counterpart 
\bea
-d/c\in[0.025,~0.030]~,
\eea
and therefore it can be neglected. 

We then compared the Coulomb to the internal nuclear action terms.
First of all we notice from Fig. \ref{fig7} that the spherical Coulomb term, plotted by open symbols,
is much larger than the nuclear one, given by dark symbols 
\bea
S^{(0)}_N \sim 2 << S^{(0)}_C\in [12,50]~. 
\eea

\begin{figure}[ht] 
\begin{center}
\includegraphics[width=8cm]{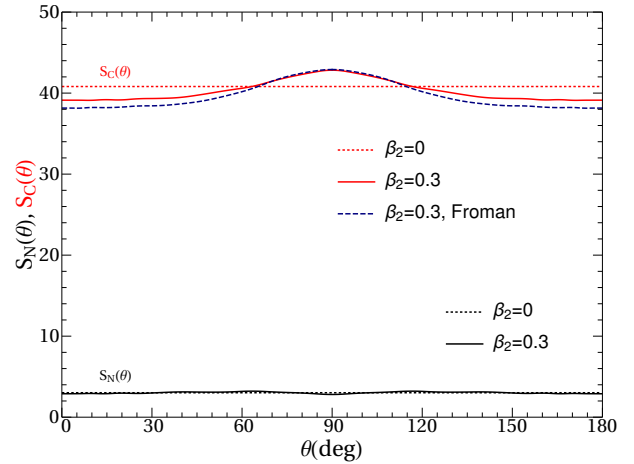} 
\caption{
Nuclear action (\ref{SNtheta}) versus $\theta$ within LWKB approach at the barrier radius $r_B$ for $\beta_2=0$ 
(dots) and for $\beta_2=0.3$ (solid line). 
The corresponding upper lines correspond to the Coulomb action (\ref{SCtheta}) within LWKB approach.
}
\label{fig8}
\end{center} 
\end{figure}

Thus, in Fig. \ref{fig8} the lower plot gives the nuclear action 
\bea
\label{SNtheta}
S_N(\theta)=S_N^{(0)}+D_N(\theta)~,
\eea
within LWKB approach at the barrier radius $r_B$ for $\beta_2=0$ by dots and for $\beta_2=0.3$ by a solid line. 
The corresponding upper lines correspond to the Coulomb action within LWKB approach.
\bea
\label{SCtheta}
S_C(\theta)=S_C^{(0)}+D_C(\theta)~,
\eea
Notice that the Fr\"oman FWKB estimate for the Coulomb action (\ref{DC}), plotted by
a dashed line, gives close values.

\begin{figure}[ht] 
\begin{center}
\includegraphics[width=9cm]{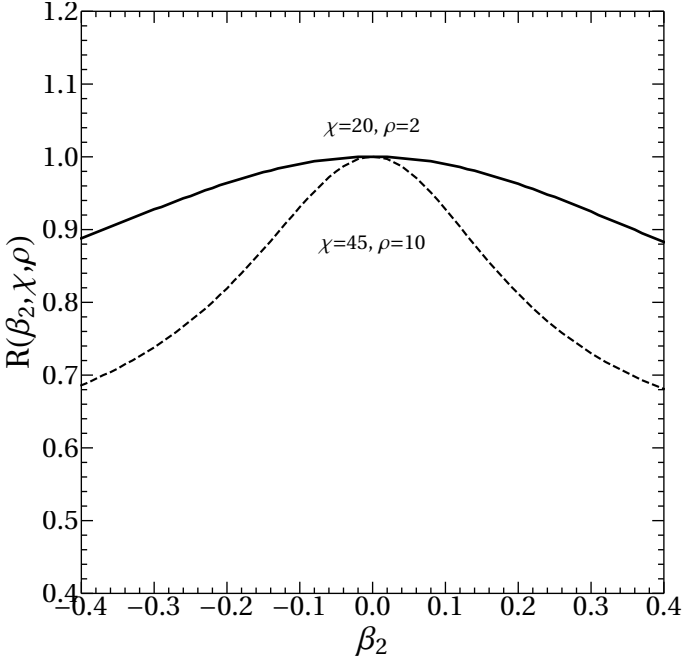} 
\caption{
Ratio between diagonal and all propagator matrix elements $R(\beta_2,\chi,\rho)$ (\ref{Dbeta}) versus quadrupole deformation 
for an alpha decay with average parameters $\chi=45,~\rho=10$ (lower dashed line) and proton emission with $\chi=20,~\rho=2$
(upper solid line).
}
\label{fig9}
\end{center} 
\end{figure}

Therefore it turns out that the angular dependence is practically given by the Coulomb terms, 
due to the fact the the nuclear part is practically constant in this scale $S_{N}(\theta)\sim S^{(0)}_N(\theta)$.
Thus, the largest internal function computing scattering amplitudes in Eq. (\ref{scater}) is
practically monopolar $|f^{(int)}_0|>>|f^{(int)}_L|,~L=2,4,...$ and therefore one obtains
the following estimate
\bea
\label{scater0}
N_L&\approx&\frac{f^{(int)}_{0}(V_{frag})}{G_L(\chi_B,\rho_B)}{\cal K}_{L0}(\beta_2,\chi_B,\rho_B)~.
\eea
Indeed, by using Eq. (\ref{fL}) with realistic channel decay widths, it turns out that 
$|f^{(int)}_L|< 10^{-3}|f^{(int)}_0|,~L=2,4,6$.
Until now we analyzed transitions between ground states. Deformation effects are probed
by the analysis of the fine-structure revealed by transitions to excited states
in the daughter nucleus.
We neglected in our formalism the contribution of the daughter dynamics. The emitted alpha-particle
with angular momentum $L$ is coupled with the same angular momentum of the daughter nucleus
to the initial spin $J_i=0$. Thus, in each channel the energy is replaced by $E\rightarrow E-E_L$,
where $E_L$ is the excitation energy of the daughter nucleus \cite{Del10}.
As we already mentioned, by neglecting non-diagonal Coriolis matrix elements in the intrinsic system of coordinates,
the decoupled system of equations at large distances (\ref{asymp}) becomes formally the same \cite{Del10,Fro57}, but
the Coulomb parameter and reduced radius for each channel are given by Eqs. (\ref{energ}).
At the barrier radius these relations become
\bea
\chi_L&=&\frac{\chi_B}{\epsilon_L}
\nn
\rho_L&=&\rho_B\epsilon_L
\nn
\epsilon_L&\equiv&\sqrt{1-\frac{E_L}{E}}~,
\eea
where the values $\chi_B,~\rho_B$ are the barrier values for $L=0$. 
Thus, the total decay width (\ref{Gamma}) becomes a superposition of channel decay widths 
as follows
\bea
\label{GammaL}
\Gamma&=&\sum_{L=even}\Gamma_L=\sum_{L=even} \hbar v_L \vert N_L\vert^2~,
\eea
in terms of the channel velocity
\bea
v_L&=&\sqrt{\frac{2E}{\mu}}~\epsilon_L~,
\eea
where the scattering amplitude (\ref{scater0}) is replaced by
\bea
\label{scaterL}
N_L&=&\frac{f^{(int)}_{0}(V_{frag}^L)}{G_L(\chi_L,\rho_L)}{\cal K}_{L0}(\beta_2,\chi_B,\rho_B)~.
\eea
Thus, each channel decay width (\ref{GammaL}) for transitions from the ground state with $J_i=0$
to final states with $J_f=L$ becomes factorized
\bea
\label{factor}
\Gamma_L=\Gamma^{(0)}_L(\chi_L,\rho_L)D_L(\beta_2,\chi_L,\rho_L)~,
\eea
into a spherical "monopole"
\bea
\label{spher}
&&\Gamma^{(0)}_L(\chi_L,\rho_L)=\hbar v_L\left[\frac{f^{(int)}_{0}(V_{frag}^L)}{G_0(\chi_L,\rho_L)}\right]^2
\nn&=&
\hbar v_Lp_0\exp\left[-2\left(S^{(0)}_C(\chi_L,\rho_L)+S_N^{(0)}(V_{frag}^L)\right)\right]~,
\nn
\eea
in terms of the channel fragmentation potential 
\bea
V_{frag}^L=V_B-(E-E_L)=V_{frag}+E_L~,
\eea
and centrifugal-deformation factor 
\bea
\label{DL}
D_L(\beta_2,\chi_L,\rho_L)&=&\exp\left[-2\frac{\tan\alpha_L}{\chi_L}L(L+1)\right]
\nn&\times&
{\cal K}^2_{L0}(\beta_2,\chi_B,\rho_B)~,
\eea
induced by the deformed Coulomb field. Here we used the exact quantum expression $L(L+1)$ due to
the fact that the alpha-decay fine structure involves low values of the angular momentum.

\begin{figure}[ht] 
\begin{center}
\includegraphics[width=9cm]{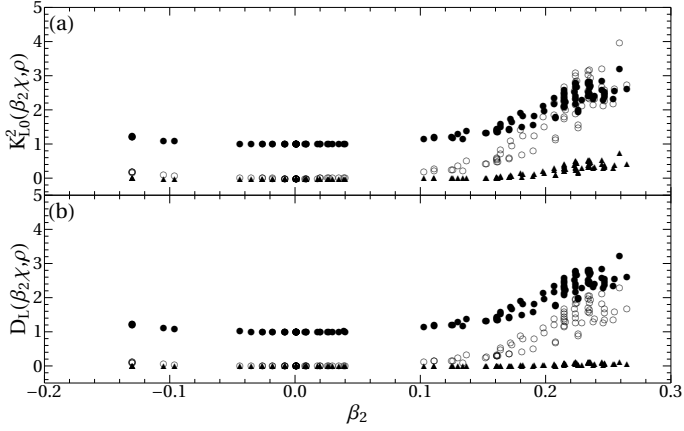} 
\caption{
(a) The deformation factor ${\cal K}^2_{L0}(\beta_2,\chi_B,\rho_B)$ versus deformation
for $L=0$ (dark circles), $L=2$ (open circles) and $L=4$ (triangles). \\
(b) Same as in (a), but for the centrifugal-deformation factor $D_{L}(\beta_2,\chi,\rho)$ (\ref{DL}).
}
\label{fig10}
\end{center} 
\end{figure}

One can also factorize the "monopole" decay width
\bea
\Gamma^{(0)}_L(\chi_L,\rho_L)&=&\gamma_0^2(V_{frag}^L)P_0(\chi_L,\rho_L)~,
\eea
in terms of the channel reduced width and penetrability
\bea
\label{factor0}
\gamma_0^2(V_{frag}^L)&=&p_0\exp\left[-2S_N^{(0)}(V_{frag}^L)\right]
\nn
P_0(\chi_L,\rho_L)&=&\hbar v_L\exp\left[-2S^{(0)}_C(\chi_L,\rho_L)\right]~,
\eea
and therefore the channel decay width can be factorized
\bea
\Gamma_L&=&\gamma_0^2(V_{frag}^L)P_L(\beta_2,\chi_L,\rho_L)~,
\eea
in terms of the channel reduced width and deformed penetrability
\bea
P_L(\beta_2,\chi_L,\rho_L)&=&P_0(\chi_L,\rho_L)D_L(\beta_2,\chi_L,\rho_L)~.
\eea
Our estimate has shown that the following approximation 
\bea
\label{DLapprox}
D_L(\beta_2,\chi_L,\rho_L)&\approx&D_L(\beta_2,\chi_B,\rho_B)~,
\eea
remains valid within 5\% accuracy at $\beta_2=0.3$ and it can be used in the above relation.
This approximate ansatz for FWKB was used in the Fr\"oman paper \cite{Fro57}, but here 
it was neglected the channel dependence of the Coulomb action and the nuclear part was not considered.
Anyway, it turns out that the main channel energy dependence connected to the daughter dynamics
is exponentially induced by Coulomb and nuclear action terms in Eq. (\ref{spher}).

Now we can explain the alignment of open symbols in Fig. \ref{fig7} along parallel straigth lines.
This feature corresponds to the well known Geiger-Nuttall law for alpha transitions between ground states
\bea
\log_{10}T_0&\sim&\log_{10}\left[\frac{G_0(r_B)}{f^{(int)}_0(r_B)}\right]^2
\sim 2\left[S^{(0)}_C+S^{(0)}_N\right] 
\nn
&=& 2\chi\left(\alpha-\oh\sin~2\alpha\right)+\frac{\pi V_{frag}}{\hbar\omega}
\nn
&\sim& a\frac{Z}{\sqrt{E}}+b~,
\eea
where $a$ and $b$ are constants.

Let us stress on the fact that the factorized representation (\ref{factor}) with
(\ref{spher}) and (\ref{DL}) remains valid for any of the above described
approximations AWKB, LWKB and FWKB. Moreover, we have shown in Fig. \ref{fig8}
that LWKB results are close to the Fr\"oman FWKB approach.
In order to point out on the deformation effect of the propagator matrix
we plotted in Fig. \ref{fig9} by the lower dashed line the following ratio
\bea
\label{Dbeta}
R(\beta_2,\chi,\rho)=\frac{S_{diag}}{S_{total}}=
\sqrt{\frac{\sum_L {\cal K}_{LL}^2(\beta_2,\chi,\rho)}{\sum_{LL'}{\cal K}_{LL'}^2(\beta_2,\chi,\rho)}}~,
\nn
\eea
for $\chi=45,~\rho=10$ by using FWKB. 
This quantity gives an overall characteristics on the coupling between channels induced by the quadrupole deformation.
One sees that the overall deformation effects are rather strong, i.e. $R\sim 0.73$
at $\beta_2\sim 0.3$. As a comparison, we plotted by the upper solid line the same ratio for proton emission
corresponding to characteristic parameters $\chi=20,~\rho=2$. Notice a significantly smaller effect
$R\sim 0.93$ at the same deformation. One can conclude that the deformation effect is mainly enhanced
by the increase of the Coulomb parameter $\chi$.

We then analyzed the influence of the deformation on each channel decay width
by plotting in Fig. \ref{fig10} (a) the deformation factor, i.e. the propagator matrix element squared
${\cal K}^2_{L0}(\beta_2,\chi_B,\rho_B)$ multiplying the spherical decay width, versus deformation
for $L=0$ (dark circles), $L=2$ (open circles) and $L=4$ (triangles).
One clearly sees that the deformation effect induced by the Coulomb barrier plays 
a significant role on each partial decay width for $\beta_2>0.1$, especially for the quadrupole $L=2$,
but also for the monopole $L=0$ channel.
In the panel (b) we plotted the centrifugal-deformation factor (\ref{DL}) versus deformation.
One clearly sees that the $L=0,~2$ channels are the most relevant in the structure of the channel decay width. 

\begin{figure}[ht] 
\begin{center}
\includegraphics[width=9cm]{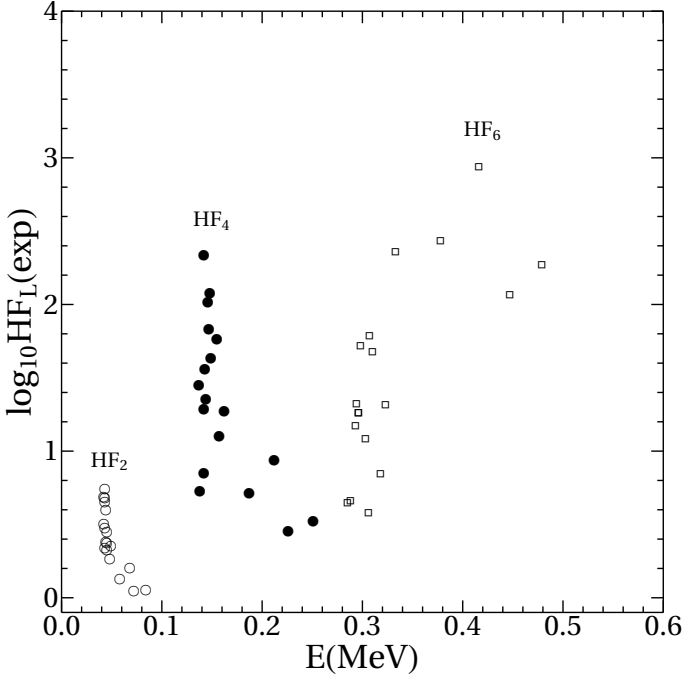} 
\caption{
Experimental hindrance factor (\ref{HFexp}) versus the excitation energy of the daughter nucleus
for $L=2$, (open circles) $L=4$ (dark circles) and $L=6$ (open squares).
}
\label{fig11}
\end{center} 
\end{figure}

The logarithm of the theoretical channel hindrance factor, estimated by using the
explicit form of the internal wave function predicts the following dependence
\bea
\label{logHF}
&&\log_{10}HF_L=\log_{10}\frac{\gamma_0^2(V_{frag})}{\gamma_0^2(V_{frag}^L)}
\nn&\sim&
\log_{10}\frac{p_0}{p_L}+2\left[S_N(V_{frag}^L)-S_N(V_{frag})\right]\log_{10}e
\nn&\sim&
\left(\log_{10}p_0-\log_{10}p_L\right)+E_L~.
\eea
We investigate the experimental hindrance factor
\bea
\label{HFexp}
HF_L(exp)=\frac{\gamma_0^2(exp)}{\gamma_L^2(exp)}~,
\eea
defined in terms of the experimental reduced width
\bea
\label{redwidL}
\gamma_L^2(exp)=\frac{\Gamma_L(exp)}{P_L(\beta_2,\chi_L,\rho_L)}~.
\eea
In Fig. \ref{fig11} we plotted $HF_L(exp)$ versus the excitation energy of the daughter nucleus $E_L$,
for $L=2$, (open circles) $L=4$ (dark circles) and $L=6$ (open squares) corresponding to
18 well deformed emitters above $^{208}$Pb with measured channel decay widths.
One indeed sees that the general trend follows the linear energy dependence of Eq. (\ref{logHF}). 
At the same time, notice the local strong decrease of each $\log_{10}HF_2$ and $\log_{10}HF_4$ 
along with the increase of the excitation energy.
This feature is given by the strong increase of the channel probability $p_L$ with respect to the 
excitation energy increase along each $L$-channel.
 
\section{Conclusions}

We compared the exact coupled channels procedure to the semiclassical approach
to describe two-body emission processes from deformed nuclei by using the propagator method.
We expressed within this approach the vector of scattering amplitudes in terms of a propagator matrix 
multiplied by the vector of internal radial wave function components 
divided to the vector of irregular Coulomb waves.
We described in a rigorous way the 3D semiclassical approach, corresponding to
deformed potentials, which leads to the exact results for the propagator matrix. 
We compared them with the much simpler expressions given by the AWKB and LWKB with its approximation, 
known as Fr\"oman method.
We have shown that LWKB approach is closer than AWKB to the exact coupled-channels formalism.
Each channel decay width becomes factorized into spherical and centrifugal-deformed terms.
An analysis of deformation effects for alpha-emission from ground states of even-even nuclei was performed.
We evidenced the important role played by deformation.

\begin{acknowledgments}
This work was supported by the grant of the Romanian 
Ministry Education and of Research PN-18090101/2019-2021 and
by the grant of the Institute of Atomic Physics from the National 
Research – Development and Innovation Plan III 
for 2015 -2020/Programme 5/Subprograme
5.1 ELI-RO, project ELI-RO No 12/2020.
\end{acknowledgments}

	
	\appendix*
	
	\section{WKB wave function \\ and quantization in 3D}
\setcounter{equation}{0} 
\renewcommand{\theequation}{A.\arabic{equation}} 

	\label{sec:Appendix}
	What we present in this appendix is not new by any means, but to the best of our knowledge, there is no comprehensive work clearly stating all considerations involved in finding a proper solution for the spherical WKB system from Eqs.~(\ref{eq:WKB_system}). We start by re-writing system (\ref{eq:WKB_system})
	\begin{align}
	\label{eq:ap_WKB_system}
	\begin{aligned}
	\hbar^0:& \left(\nabla S_0^{(0)}(\textbf{r})\right)^2 = -K_0^2(r)\\
	\hbar^1:& -\frac{i}{2}\Delta S_0^{(0)}(\textbf{r}) + (\nabla S_0^{(0)}(\textbf{r}))(\nabla S_0^{(1)}(\textbf{r})) = 0
	\end{aligned}
	\end{align}
	
	The equation for the first order in $\hbar$ can be solved by means of separation of variables.
	We, thus, write
	\begin{equation*}
	S^{(0)}_{0}(\textbf{r}) = A^{(0)}(r) + B^{(0)}(\theta) + C^{(0)}(\phi)
	\end{equation*}
	and obtain
	\begin{align}
	\label{eq:ap_WKB_spher_separated}
	\begin{aligned}
	&\left[\nabla S_{0}^{(0)}(\textbf{r})\right]^2 = \\
	&=\left(\frac{dA^{(0)}(r)}{dr}\right)^2 + \left(\frac{1}{r}\frac{dB^{(0)}(\theta)}{d\theta}\right)^2 +
	\left(\frac{1}{r\sin\theta} \frac{dC^{(0)}(\phi)}{d\phi}\right)^2 \\
	&=-2\mu\left(V_0(r)-E\right)
	\end{aligned}
	\end{align}
	from here on we employ the notation
	\begin{equation}
	\label{eq:ap_Kspher_def}
	K_{0}(r)\equiv \sqrt{2\mu E\left[\frac{V_0(r)}{E}-1\right]}
	\end{equation}
	We separate $\phi$ in the above equation and solve for $C(\phi)$
	\begin{align}
	\begin{aligned}
	&-\left(\frac{dC^{(0)}}{d\phi}\right)^2 = -\lambda_{\phi}^2 \\
	&=r^2\sin^2\theta\left[\left(\frac{dA^{(0)}}{dr}\right)^2 
	+ \frac{1}{r^2}\left(\frac{dB^{(0)}}{d\theta}\right)^2 + K^2_0\right]
	\end{aligned}
	\end{align}
	where $\lambda_{\phi}$ is a separation constant, which will be determined through
	quantization. As expected, we obtain a periodic dependence on $\phi$ in our wave function 
	through
	\begin{equation}
	\label{eq:ap_Cshper_form}
	C^{(0)}(\phi) = \pm\lambda_{\phi}\phi
	\end{equation}
	We now separate $\theta$
	\begin{align}
	\label{eq:ap_B0_diffEq}	
	\begin{aligned}
	-\left(\frac{dB^{(0)}}{d\theta}\right)^2 - \frac{\lambda_{\phi}^2}{\sin^2\theta} &= -\lambda_{\theta}^2\\
	&=r^2\left[\left(\frac{dA^{(0)}}{dr}\right)^2+K_0^2\right]
	\end{aligned}
	\end{align}
	where $\lambda_{\theta}$ is another separation constant. Rearranging the $\theta$ part gives
	\begin{equation}
	\frac{dB^{(0)}}{d\theta} = \pm\sqrt{\lambda_{\theta}^2 - \frac{\lambda_{\phi}^2}{\sin^2\theta}}
	\end{equation}
	The closed form of $B(\theta)$ is not,
	at this point, of interest to us so we proceed with the last variable for which
	\begin{equation}
	\left(\frac{dA^{(0)}}{dr}\right)^2 = -K_0^2 - \frac{\lambda_{\theta}^2}{r^2}
	\end{equation}
	which, upon expanding all terms, takes the familiar form
	\begin{equation}
	\frac{dA^{(0)}}{dr} = \pm i\sqrt{2\mu E\left(\frac{V_{0}(r)}{E} - 1 + \frac{\lambda_{\theta}^2}{r^2}\right)}
	\end{equation}
	Now we could address the quantization procedure. The astute reader can already guess that 
	if one stops here and performs the quantization only to the first order, 
	the separation constants would become \cite{Cur04} (by straight forward identification)
	\begin{align}
	\label{eq:ap_separation_consts}
	\begin{aligned}
	\lambda_{\phi} &= M\hbar\\
	\lambda_{\theta} &= \left(L+\frac{1}{2}\right)\hbar
	\end{aligned}
	\end{align}
	where $M$ is the magnetic quantum number and $L$ is the usual orbital quantum number. A full account of the above expressions and the reason for the Langer correction ($\sqrt{L(L+1)} \to L+1/2$) \cite{Lan37} will be given later on. However, to increase the accuracy of the WKB method we have to compute also the second order contribution which, rigorously speaking, must enter in the quantization procedure. We notice that the second equation in (\ref{eq:ap_WKB_system}) is also separable given the expression we found for the first order contribution. We can thus write
	\begin{equation}
	S_0^{(1)}(\textbf{r}) = A^{(1)}(r)+B^{(1)}(\theta) +C^{(1)}(\phi)  
	\end{equation}
	which implies (by direct substitution)
	\begin{align}
	\label{eq:ap_WKB_spher_second_order}
	\begin{aligned}
	&\frac{i}{2}\left(\frac{d^2 A^{(0)}}{dr^2} + \frac{2}{r} \frac{d A^{(0)}}{dr}
	+\frac{1}{r^2}\frac{d^2B^{(0)}}{d\theta^2} + \frac{\cot\theta}{r^2}\frac{dB^{(0)}}{d\theta}\right)\\
	=& \frac{dA^{(0)}}{dr}\frac{dA^{(1)}}{dr} + \frac{1}{r^2}\frac{dB^{(0)}}{d\theta}\frac{dB^{(1)}}{d\theta} 
	+\frac{1}{r^2\sin^2\theta}\frac{dC^{(0)}}{d\phi}\frac{dC^{(1)}}{d\phi} 
	\end{aligned}
	\end{align}
	We separate first the $\phi$ dependence and obtain
	\begin{align}
		\begin{aligned}
		&\frac{i r^2\sin^2\theta}{2}\left(\frac{d^2 A^{(0)}}{dr^2} + \frac{2}{r} \frac{d A^{(0)}}{dr}
		+\frac{1}{r^2}\frac{d^2B^{(0)}}{d\theta^2} + \frac{\cot\theta}{r^2}\frac{dB^{(0)}}{d\theta}\right)\\
		&-r^2\sin^2\theta\left(\frac{dA^{(0)}}{dr}\frac{dA^{(1)}}{dr} + \frac{1}{r^2}\frac{dB^{(0)}}{d\theta}\frac{dB^{(1)}}{d\theta} \right)\\
		&= \frac{dC^{(0)}}{d\phi}\frac{dC^{(1)}}{d\phi} = \gamma_{\phi}
		\end{aligned}
	\end{align}
	where $ \gamma_{\phi} $ is another separation constant and, solving for $C^{(1)}$ we find
	\begin{equation}
	C^{(1)}(\phi) = \pm \frac{\gamma_{\phi}}{\lambda_{\phi}}\phi.
	\end{equation}
	where $\gamma_{\phi}$ is another separation constant. We address now the quantization of the $ \phi $ motion. The generally accepted semiclassical quantization is the Einstein-Brillouin-Keller (EBK) condition which reads \cite{Bra03}
	\begin{equation}
		\label{eq:ap_EBK_quantization}
		\frac{1}{2\pi}\oint P_{q}dq = \left(n_{q}+\frac{\mu_i}{4} + \frac{b_i}{2}\right)\hbar
	\end{equation}
	where $q$ is the generalized variable, $P_{q}$ is its associated generalized momentum, $n_{q}$ is the standard quantum number for that variable, $\mu_q, b_q$ are Maslov indexes ($\mu_q$ is the number of conventional turning points along the integration path and $b_{q} $ is the number of hard-wall turning points along the integration path). The integration path is the path traversed by the classical particle in one complete period. In the case of the $\phi$ motion, the integration path is $[0,2\pi]$ since this corresponds to a complete $\phi$ period and the whole range is classically allowed. The generalized momentum is given by
	\begin{equation}
		\label{eq:ap_phi_gen_mom}
		P_{\phi}\equiv\frac{\partial S}{\partial \phi} = \left(\lambda_{\phi} + \hbar \frac{\gamma_{\phi}}{\lambda_{\phi}}\right)
	\end{equation}
	The quantization condition reads
	\begin{equation}
	\frac{1}{2\pi}\oint P_{\phi}d\phi = \frac{1}{2\pi}\int_{0}^{2\pi}d\phi\left(\lambda_{\phi} + \hbar \frac{\gamma_{\phi}}{\lambda_{\phi}}\right) = n_{\phi}\hbar
	\end{equation}
	since the integrand and we obtain
	\begin{equation}
	\label{eq:ap_lambda_phi_quant}
	\lambda_{\phi} +\hbar\frac{\gamma_{\phi}}{\lambda_{\phi}} =  n_{\phi}\hbar
	\end{equation}
	
	We now turn back to $\theta$ and apply a similar reasoning as we did for $\phi$
	\begin{align}
		\begin{aligned}
		&\frac{dB^{(0)}}{d\theta} \frac{dB^{(0)}}{d\theta} + \frac{\gamma_{\phi}}{\sin^2\theta}-\frac{i}{2}\frac{d^2 B^{(0)}}{d\theta^2} - \frac{i\cot\theta}{2}\frac{dB^{(0)}}{d\theta}\\
		&=\frac{i r^2}{2}\left(\frac{d^2 A^{(0)}}{dr^2} + \frac{2}{r} \frac{d A^{(0)}}{dr}\right)\\
		&=\gamma_{\theta}
		\end{aligned}
	\end{align}
	where $\gamma_{\theta}$ is another separation constant. After some rearrangements we can write
	\begin{align}
	\begin{aligned}
	\frac{dB^{(1)}}{d\theta} &= \frac{i}{2}\frac{1}{\frac{dB^{(0)}}{d\theta}}\frac{d^2B^{(0)}}{d\theta^2} 
	+ \frac{i}{2}\cot\theta + \\
	&\frac{1}{\frac{dB^{(0)}}{d\theta}}\left(\gamma_{\theta} -
	\frac{\gamma_{\phi}}{\sin^2\theta}\right)
	\end{aligned}
	\end{align}
	with $\gamma_{\theta}$ another separation constant. A discussion is called for here regarding the last two separation constants. With the constraints derived up to now, they could take any value subject, of course, to the quantization conditions. We saw that for the $\phi$ dependence, $\gamma_{\phi}$ does not make any difference aside from a phase. The problem, however arises for the $\theta$ dependence. If we now try to make the analogy with the exact result, we see that $\gamma_{\phi}$ and $\gamma_{\theta}$ should be set to 0. Indeed starting from the system of equations for the Legendre associated functions $P_{L,M}(\theta)$ and the azimuth function $\Phi$
	\begin{align*}
		\frac{d^2 P_{L,M}}{d\theta^2} +& \cot\theta \frac{dP_{L,M}}{d\theta} + \\
		&\left(L(L+1) - \frac{M^2}{\sin^2\theta}\right)P_{L,M} = 0\\
		&\frac{d^2 \Phi}{d\phi^2} = -M^2\Phi
	\end{align*}
	and perform the semiclassical expansion on both equations independently, we see that only 2 constants arise. This implies that $\gamma_{\phi} = \gamma_{\theta} = 0$.
	
	In the light of the above considerations we can write
	\begin{equation}
		\frac{dB^{(1)}}{d\theta} = \frac{i}{2}\frac{1}{\frac{dB^{(0)}}{d\theta}}\frac{d^2B^{(0)}}{d\theta^2} 
		+ \frac{i}{2}\cot\theta
	\end{equation}
	with the solution given by
	\begin{equation}
		B^{(1)}(\theta) = \frac{i}{2}\log\left|\frac{dB^{(0)}}{d\theta}\right| + \frac{i}{2}\log\sin(\theta)
	\end{equation}
	
	Now we have to determine $\lambda_{\theta}$ which is done through the quantization condition
	\begin{equation}
		\label{eq:ap_Theta_quantization}
		\frac{1}{2\pi}\oint P_{\theta}d\theta = \left(n_{\theta} + \frac{\mu_{\theta}}{4} + \frac{b_{\theta}}{2}\right)\hbar
	\end{equation}
	with 
	\begin{equation}
		P_{\theta} \equiv \frac{dS_0}{d\theta} = \frac{dB^{(0)}}{d\theta} + \hbar \frac{dB^{(1)}}{d\theta}
	\end{equation}
	In this case, the classically allowed range for $\theta$ is $[\frac{\pi}{2} - \gamma, \frac{\pi}{2} - \gamma]$ where $\gamma = \arccos(\lambda_{\phi}/\lambda_{\theta})$, meaning that the integration contour is 2 times this range. Moreover, there are two classical turning points at the end of the range with no hard-walls, hence $\mu_{\theta} = 2$ and $b_{\theta} = 0$ so Eq.~(\ref{eq:ap_Theta_quantization}) becomes
	\begin{equation}
		\label{eq:ap_Quantization_Theta}
		\frac{1}{\pi}\int_{\frac{\pi}{2}-\gamma}^{\frac{\pi}{2}+\gamma}d\theta \sqrt{\lambda_{\theta}^2 - \frac{\lambda_{\phi}^2}{\sin^2\theta}} \equiv I_{\theta} = \left(n_{\theta} +\frac{1}{2}\right)\hbar
	\end{equation}
	because the logarithms evaluated along this contour give no contribution (no poles inside the integration domain). To solve the integral above, we follow the approach in chapter 13 of \cite{Gol02}, but we mention that the $\theta$ allowed region is the one given above. First we change the variable using
	\begin{equation*}
		\cos\theta=\sin\gamma\sin\eta
	\end{equation*}
	which gives after some rearrangements
	\begin{equation*}
		I_{\theta} = 
		\frac{\lambda_{\theta}}{\pi}\int_{-\frac{\pi}{2}}^{\frac{\pi}{2}}d\eta \frac{\sin^2\gamma \cos^2\eta}{1-\sin^2\gamma\sin^2\eta}
	\end{equation*}
	then, with another change of variable
	\begin{equation*}
		u=\tan\eta
	\end{equation*}
	the integral becomes
	\begin{align}
		\label{eq:ap_Lambda_Theta_Quant}
		I_{\theta} &= \frac{\lambda_{\theta}}{\pi} \int_{-\infty}^{\infty}du\frac{\sin^2\gamma}{\left(1+u^2\right)\left(1+u^2\cos^2\gamma\right)}\\
		&=\frac{\lambda_{\theta}}{\pi}\int_{-\infty}^{\infty} du\left(\frac{1}{1+u^2} - \frac{\cos^2\gamma}{1+u^2\cos^2\gamma}\right)\\
		&=\frac{\lambda_{\theta}}{\pi}\left(\arctan u - \cos\gamma\arctan\left(u\cos\gamma \right)\right)\Big|_{-\infty}^{\infty}\\
		&=\lambda_{\theta}-\lambda_{\phi}
	\end{align}
	Adding the result above to Eq.~(\ref{eq:ap_lambda_phi_quant}) and taking into account that $\gamma_{\phi} = 0$ gives
	\begin{equation}
		\lambda_{\theta} = \left(n_{\theta} + n_{\phi} + \frac{1}{2}\right)\hbar
	\end{equation}
	which, if we denote $L = n_{\theta} + n_{\phi}$, can be written as
	\begin{equation}
		\lambda_{\theta} = L+\frac{1}{2}
	\end{equation}
	
	Now, we can solve for $r$ and we find that
	\begin{equation*}
	\frac{dA^{(1)}}{dr} = \frac{i}{2}\frac{1}{\frac{dA^(0)}{dr}}\frac{d^2A^{(0)}}{dr^2} + ir
	\end{equation*}
	which gives
	\begin{equation}
	A^{(1)}(r) = \frac{i}{2}\log\left(\left|\frac{dA^{(0)}}{dr}\right|\right)+i\log(r)
	\end{equation}
	We now gather all results together and obtain the WKB approximation of the wave function
	for the 3D motion as a superposition of "incoming" and "outgoing" functions
\begin{widetext}
	\begin{align}
	\label{eq:ap_WKB_fullWf_spherical}
	\begin{aligned}
	\Psi_{0}(\textbf{r}) = \frac{Y^{(\textrm{WKB})}_{LM}(\theta,\phi)} {\sqrt{K_{0,L}(r)}}
	\left\lbrace c^{(out)}_{L} \exp\left[\int_{r_{0}}^{r}dr' K_{0,L}(r')\right] + 
	c^{(in)}_{L} \exp\left[-\int_{r_{0}}^{r}dr'K_{0,L}(r')\right]\right\rbrace~,
	\end{aligned}
	\end{align}
\end{widetext}
	where $Y^{(\textrm{WKB})}_{L,M}$ is the WKB approximation of the spherical harmonic $Y_{L,M}$ and $r_{0}$ is the starting point of integration. We do not give the closed form of $Y_{L,M}$ since it is more complicated and not useful as we can use the exact result. However, the reader is advised to consult ref. \cite{Mor91} for a complete account, or \cite{Lan77}, for the case $M=0$. We also mention here the works of Robnik \cite{Rob97,Rob97a} who attempts a general quantization to all orders under some conjecture and the work of Salasnich \cite{Sal97}. Both authors show that under some special circumstances, the quantum eigenvalue of the angular momentum operator can be retrieved from semiclassical calculations. 
	
	Finally, since it is helpful for the studies in this work, we give here the form of the radial part of an outgoing solution of the spherical problem
	\begin{align}
	\label{eq:ap_GL_defintion}
	\begin{aligned}
	G_{L}(r) &= \frac{1}{\sqrt{K_{0,L}(r)}}\exp\left[\int_{r}^{r_{2,L}} dr'K_{0,L}(r')\right]~,
	\end{aligned}
	\end{align}
	where $r_{2,L}$ is the external turning point.


\end{document}